# An Explicit Construction of 2-Generator Quasi-Twisted Codes

Eric Z. Chen

*Abstract*— Quasi-twisted (QT) codes are a generalization of quasi-cyclic (QC) codes. Based on consta-cyclic simplex codes, a new explicit construction of a family of 2-generator quasi-twisted (QT) two-weight codes is presented. It is also shown that many codes in the family meet the Griesmer bound and therefore are length-optimal. New distance-optimal binary QC [195, 8, 96], [210, 8, 104] and [240, 8, 120] codes, and good ternary QC [208, 6, 135] and [221, 6, 144] codes are also obtained by the construction.

## I. INTRODUCTION

AS a generalization to cyclic codes, quasi-cyclic (QC) codes and quasi-twisted (QT) codes have been shown to contain many good linear codes. Many researchers have been using modern computers to search for good QC or QT codes, and many record-breaking codes are found [1]–[12]. The problem with this method is that it becomes intractable when the dimension and the length of the code become large. Unfortunately, very little is known on explicit constructions of good QC or QT codes. For *2*-generator QC or QT codes, even fewer results are known [13], [14].

A linear code is called projective if any two of its coordinates are linearly independent, or in other words, if the minimum distance of its dual code is at least three. A code is said to be two-weight if it has only two non-zero weights. Projective two-weight codes are closely related to strongly regular graphs [15].

In this paper, a new explicit construction of a family of *2*-generator QT two-weight codes is presented. It is the first time that a family of 2-generator QT codes is constructed systematically. It is also shown that many codes of this family are good and optimal. Examples are given to show the construction and the modular structure of the codes.

## II. QUASI-TWISTED CODES AND TWO-WEIGHT CODES

### A. Consta-Cyclic Codes

The codes discussed in the following sections are linear. A q-ary linear code is a k-dimensional subspace of an n-dimensional vector space over the finite field $F_q$, with minimum distance d between any two codewords. We denote a q-ary code as an $[n, k, d]_q$ code, or a binary $[n, k, d]$ code if q = 2. A linear $[n, k, d]_q$ code is said to be $\lambda$-consta-cyclic if there is a non-zero element $\lambda$ of $F_q$ such that for any codeword $(a_0, a_1, ..., a_{n-1})$, a consta-cyclic shift by one position or $(\lambda a_{n-1}, a_0, ..., a_{n-2})$ is also a codeword [16]. Therefore, the consta-cyclic code is a generalization of the cyclic code, and a cyclic code is a $\lambda$-consta-cyclic code with $\lambda = 1$. A consta-cyclic code can be defined by a generator polynomial.

### B. Hamming Codes and Simplex Codes

Hamming codes are a family of linear single error correcting codes. For any positive integer t > 1 and prime power q, we have a Hamming $[n, n–t, 3]_q$ code, where $n = (q^t-1)/(q-1)$. Further, if t and q–1 are relatively prime, then the Hamming code is equivalent to a cyclic code.

The dual code of a Hamming code is called the simplex code. So for any integer t > 1 and prime power q, there is a simplex $[(q^t-1)/(q-1), t, q^{t-1}]_q$ code. It should be noted that a simplex code is an equidistance code, where $q^t-1$ non-zero codewords have a weight of $q^{t-1}$. Let h(x) be a primitive polynomial of degree *t* over $F_q$. A $\lambda$-consta-cyclic simplex $[(q^t-1)/(q-1), t, q^{t-1}]_q$ code can be defined by the generator polynomial $g(x) = (x^n-\lambda)/h(x)$, where $n = (q^t-1)/(q-1)$ and $\lambda$ has order of q–1 [16]. Further, a simplex code is equivalent to a cyclic code if t and q–1 are relatively prime.

### C. Quasi-Twisted Codes

A code is said to be quasi-twisted (QT) if a consta-cyclic shift of any codeword by p positions is still a codeword. Thus a consta-cyclic code is a QT code with p = 1, and a quasi-cyclic (QC) code is a QT code with $\lambda = 1$. The length n of a QT code is a multiple of p, i.e., n = pm.

The consta-cyclic matrices are also called twistulant matrices. They are basic components in the generator matrix for a QT code. An m × m consta-cyclic matrix is defined as

$$C = \begin{bmatrix} c_0 & c_1 & c_2 & \cdots & c_{m-1} \\ \lambda c_{m-1} & c_0 & c_1 & \cdots & c_{m-2} \\ \lambda c_{m-2} & \lambda c_{m-1} & c_0 & \cdots & c_{m-3} \\ \vdots & \vdots & \vdots & \vdots & \vdots \\ \lambda c_1 & \lambda c_2 & \lambda c_3 & \cdots & c_0 \end{bmatrix}, \quad (1)$$

and the algebra of m × m consta-cyclic matrices over $F_q$ is isomorphic to the algebra in the ring $f[x]/(x^m-\lambda)$ if C is mapped onto the polynomial formed by the elements of its first row, $c(x) = c_0 + c_1 x + ... + c_{m-1} x^{m-1}$, with the least significant coefficient on the left. The polynomial c(x) is also called the defining polynomial of the matrix C. A twistulant matrix is called a circulant matrix if $\lambda = 1$.

The generator matrix of a QT code can be transformed into rows of twistulant matrices by suitable permutation of

Eric Z. Chen is with Dept. of Computer Science, Kristianstad University College, 291 88 Kristianstad, Sweden( eric.chen@hkr.se).

columns. So a 1-generator QT code has the following form of the generator matrix [17]:

$$G = [\, G_0 \; G_1 \; G_2 \; \ldots \; G_{p-1} \,], \quad (2)$$

where $G_i$, $i = 0, 1, \ldots, p-1$, are twistulant matrices of order m. A 2-generator QT code has generator matrix of the following form:

$$G = \begin{bmatrix} G_{00} & G_{01} & \ldots & G_{0,p-1} \\ G_{10} & G_{11} & \ldots & G_{1,p-1} \end{bmatrix}, \quad (3)$$

where $G_{ij}$ are twistulant matrices, for $i = 0, 1$, and $j = 0, 1, \ldots, p-1$.

Very little on 2-generator QT codes is known in the literature. Two 2-generator QC codes were given in [13], while a construction method for 2-generator QC codes was presented in [14]. In both cases, the codes were constructed by the help of modern computers. In the following section, we will present an explicit construction of a family of 2-generator QT codes, that are also two-weight codes, and we will prove that for many parameters they are good or optimal.

### D. Two-Weight Codes

Let $w_1$ and $w_2$, be the two non-zero weights of a two-weight code, where $w_1 \neq w_2$. We denote a projective q-ary linear two weight code as an $[n, k; w_1, w_2]_q$ code. There is an online database of two-weight codes [20]. In the survey paper [15], Calderbank and Kantor listed many known families of projective two-weight codes. Among those families, there is a family of two-weight $[n, k; w_1, w_2]_q$ codes, denoted by SU2. For any prime power q, positive integers $t > 1$ and i, the codes have the following parameters:

Block length: $n = i(q^t-1)/(q-1)$,
Dimension: $k = 2t$,
Weights: $w_1 = (i-1) q^{t-1}$, $w_2 = iq^{t-1}$,

where $2 \leq i \leq q^t$. No explicit construction of the code is known and studied in the literature, and very little is known about its structure and performance.

### III. 2-GENERATOR QUASI-TWISTED TWO-WEIGHT CODES

#### A. Explicit Construction

For any integer $t > 1$ and prime power q, a $\lambda$-consta-cyclic simplex $[(q^t-1)/(q-1), t, q^{t-1}]_q$ code can be constructed. Let $g(x)$ be the generator polynomial for the $\lambda$-consta-cyclic simplex code. Let $a_1, a_2, \ldots, a_{q-1}$ be $q-1$ non-zero elements of $F_q$ and let $m = (q^t-1)/(q-1)$. Any codeword of the $\lambda$-consta-cyclic simplex $[(q^t-1)/(q-1), t, q^{t-1}]_q$ code can be expressed by a polynomial $g_{i,j}(x) = a_i x^j g(x)$, with the computation modulo $x^m - \lambda$, where $i = 1, 2, \ldots, q-1$, and $j = 0, 1, \ldots, m-1$. Let $G_t$ be the twistulant matrix defined by the generator polynomial $g(x)$, and $G_{i,j}$ be the twistulant matrix defined by $g_{i,j}(x)$. So totally, we obtain $q^t - 1$ twistulant matrices. We construct a generator matrix of a 2-generator QT code as follows:

$$G = \begin{bmatrix} G_t & G_t & \ldots & G_t \\ 0 & B_1 & \ldots & B_{p-1} \end{bmatrix} = \begin{bmatrix} G_1 \\ G_2 \end{bmatrix}, \quad (4)$$

where $B_1, \ldots, B_{p-1}$ are $p-1$ different twistulant matrices from the $q^t-1$ twistulant matrices $\{ G_{i,j} \mid i = 1, 2, \ldots, q-1$ and $j = 0, 1, \ldots, m-1\}$ defined above, $G_1$ and $G_2$ are the first and second row of the twistulant matrices of G. Then we have following result:

**Theorem 1:** For any positive integer $t > 1$, and prime power q, the generator matrix given in (4) defines a 2-generator QT two-weight $[pm, 2t; (p-1)q^{t-1}, pq^{t-1}]_q$ code, where $m = (q^t-1)/(q-1)$ and $p = 2, 3, \ldots, q^t$.

**Proof**: A $\lambda$-consta-cyclic simplex $[(q^t-1)/(q-1), t, q^{t-1}]_q$ code is an equidistance code. Its $q^t-1$ nonzero codewords can be partitioned into $q-1$ groups of codewords, such that each group has $m = (q^t-1)/(q-1)$ codewords, and these codewords are $\lambda$-consta-cyclic each other. Let m(x) be the message polynomial to be encoded by a defining polynomial $g_{i,j}(x) = a_i x^j g(x)$. Let $b_1(x)$ and $b_2(x)$ be two polynomials corresponding to the distinct twistulant matrices $B_1$ and $B_2$. Then the codewords obtained by encoding m(x) by $b_1(x)$ and $b_2(x)$, i.e., $m(x)b_1(x)$ and $m(x)b_2(x)$, will be different, where the multiplication is computed with modulo $x^m - \lambda$. Let $C_1$ be the sub-code defined by $G_1$, and $C_2$ be the sub-code defined by $G_2$. So $C_1$ consists of codewords that just repeat the codewords of the consta-cyclic simplex $[m, t, q^{t-1}]_q$ code by p times. Therefore, $C_1$ is also an equidistance code with a distance of $pq^{t-1}$. Similarly, $C_2$ is also an equidistance code with a distance of $(p-1)q^{t-1}$, since its codeword consists of first m zeros, followed by $(p-1)$ different codewords of the consta-cyclic simplex $[m, t, q^{t-1}]_q$ code. Based on the equidistance property of the simplex code and the generator matrix structure of (4), the sum of non-zero codewords from $C_1$ and $C_2$ has a weight of $(p-1)q^{t-1}$, or $pq^{t-1}$, depending on the codewords in $C_1$ and $C_2$ have the same codeword from the consta-cyclic simplex $[m, t, q^{t-1}]_q$ code or not. Therefore, any non-zero codeword of the 2-generator QC code defined by (4) has a weight $w_1 = (p-1)q^{t-1}$ or $w_2 = pq^{t-1}$. This proves Theorem 1. Q.E.D.

Theorem 1 provides the complete solution to this family of two-weight codes. The results presented in [21] are two special cases. We state them as corollaries. When q = 2, the binary simplex $[2^t-1, t, 2^{t-1}]$ code is cyclic. So the constructed two-weight code is quasi-cyclic.

**Corollary 2:** Let q = 2. For any positive integer $t > 1$, the generator matrix (4) defines a 2-generator QC two-weight $[p(2^t-1), 2t; (p-1)2^{t-1}, p2^{t-1}]$ code, where $p = 2, 3, \ldots, 2^t$.

If t and $q-1$ are relatively prime, the simplex $[(q^t-1)/(q-1), t, q^{t-1}]_q$ code is also a cyclic code, and thus we have the following result:

**Corollary 3:** Let $t > 1$ be any positive integer, and q be a prime power. If t and $q-1$ are relatively prime, the generator matrix (4) defines a 2-generator QC two-weight $[pm, 2t; (p-1)q^{t-1}, pq^{t-1}]_q$ code, where $m = (q^t-1)/(q-1)$ and $p = 2, 3, \ldots, q^t$.

#### B. Examples

Example 1: Let q = 2 and t = 3. Then we have $x^7 - 1 = (x+1)(x^3 + x + 1)(x^3 + x^2 + 1)$. So a binary cyclic simplex [7, 3, 4] code can be defined by $g(x) = x^4 + x^2 + x + 1$. Let $G_3$ be the circulant matrix defined by the generator polynomial g(x), and $G_{1,j}$ be the circulant matrices defined by $g_{1,j}(x) = x^j g(x)$, where $j = 0, 1, 2, \ldots, 6$. Then the following generator matrix defines a binary 2-generator QC two-weight [56, 6; 28, 32] code:

$$G = \begin{bmatrix} G_3 & G_3 & G_3 & G_3 & G_3 & G_3 & G_3 & G_3 \\ 0 & G_{1,0} & G_{1,1} & G_{1,2} & G_{1,3} & G_{1,4} & G_{1,5} & G_{1,6} \end{bmatrix},$$

If we take p columns of the matrices in the above generator matrix, we obtain other binary 2-generator QC two-weight [14, 6; 4, 8], [21, 6; 8, 12], [28, 6; 12, 16], [35, 6; 16, 20], [42, 6; 20, 24], and [49, 6; 24, 28] codes in the series, for $1 < p < 8$.

Example 2: Let $q = 3$ and $t = 3$. Then $m = 13$ and $q-1 = 2$. Since 3 and 2 are relatively prime, we have a cyclic simplex $[13, 3, 9]_3$ code. It can be shown that $g(x) = x^{10} - x^9 + x^8 - x^6 - x^5 + x^4 + x^3 + x^2 + 1$ defines a cyclic simplex $[13, 3, 9]_3$ code. So a series of 2-generator QC two-weight $[13p, 6; 9(p-1), 9p]_3$ can be constructed, with $p = 2, 3, \ldots, 27$.

Example 3: Let $q = 3$ and $t = 2$. Then $m = 4$, $q-1 = 2$, and $\lambda = 2$. Since t and $q - 1$ are not relatively prime, we have a 2-consta-cyclic $[4, 2, 3]_3$ code. One primitive polynomial of degree 2 over $F_3$ is $h(x) = x^2 - x - 1$. So the corresponding generator polynomial for this 2-consta-cyclic code is $g(x) = (x^4 - 2)/h(x) = x^2 + x - 1$. Let $a_1 = 1$, $a_2 = 2$. Let $G_2$ be the twistulant matrix defined by $g(x)$, and $G_{i,j}$ be the twistulant matrices defined by $a_i x^j g(x)$, for $i = 1, 2$, and $j = 0, 1, 2$ and 3. With Theorem 1, a series of 2-generator QT two-weight $[4p, 4; 3(p-1), 3p]_3$ codes can be constructed for $p = 2, 3, \ldots, 9$.

## IV. GOOD AND OPTIMAL CODES

### A. Distance-Optimal Codes

A linear $[n, k, d]_q$ code is distance-optimal or d-optimal if its minimum distance cannot be improved, i.e., if there does not exist an $[n, k, d+1]_q$ code. For given n, k and q, there is an online table of best-known linear codes [18]. It provides the bound on the minimum distance of a code. For some parameters, the bounds are exact, while for others, the lower and upper bounds are presented. A code that meets the bound is d-optimal. A code is said to be good if it reaches the lower bound on the minimum distance, since no codes with larger distance are known. For binary quasi-cyclic codes, there is an online database on best-known binary QC codes [19]. Among the 2-generator QC and QT two-weight codes constructed above, many codes are good, and d-optimal. For example, the binary 2-generator QC two-weight $[7p, 6; 4(p-1), 4p]$ codes with $2 < p \leq 8$, and $[15p, 8; 8(p-1), 8p]$ codes with $9 < p \leq 16$ are d-optimal. The 2-generator QT two-weight $[4p, 4; 3(p-1), 3p]_3$ code with $2 < p \leq 9$, the 2-generator QC $[5p, 4; 4(p-1), 4p]_4$ code with $6 < p \leq 16$, and the 2-generator QT $[6p, 4; 5(p-1), 5p]_5$ code with $12 < p \leq 25$ are d-optimal too. Among these codes, [195, 8, 96], [210, 8, 104] and [240, 8, 120] codes are previously unknown to be quasi-cyclic [19]. By computer search for good codes, Gulliver and Bhargava constructed 1-generator QC $[36, 4, 23]_3$ code [22]. But with $q = 3$, $t = 2$, and $p = 9$, a d-optimal 2-generator QT $[36, 4, 24]_3$ code can be constructed by our method. With $q = 3$, $t = 3$, and $p = 16$ and $p = 17$, 2-generator QC two-weight $[208, 6; 135, 144]_3$ and $[221, 6; 144, 153]_3$ codes can be obtained and they reach the lower bound on the distance [18].

### B. Length-Optimal Codes

A linear $[n, k, d]_q$ code is length-optimal or n-optimal if its length cannot be improved, i.e., if there does not exist an $[n-1, k, d]_q$ code. For an $[n, k, d]_q$ code, the block length n is ruled by Griesmer bound [16]:

$$n \geq \sum_{j=0}^{k-1} \left\lceil \frac{d}{q^j} \right\rceil, \quad (5)$$

where $\lceil x \rceil$ denotes the smallest integer greater than or equal to x.

Let $p = q^t$. Then a simplex code with dimension 2t can be constructed in a 2-generator QT form:

**Theorem 4**: For any prime power q and integer $t > 1$, let $p = q^t$ and $m = (q^t-1)/(q-1)$. The following generator matrix defines a 2-generator QT simplex $[(q^{2t}-1)/(q-1) = (p+1)m, 2t, q^{2t-1}]_q$ code:

$$G = \begin{bmatrix} G_t & G_t & \ldots & G_t & 0 \\ 0 & B_1 & \ldots & B_{p-1} & G_t \end{bmatrix}. \quad (6)$$

This theorem can be proved similar to Theorem 1. So the proof is omitted.

The given examples show that many SU2 codes are good or distance-optimal. The following theorem tells how good the codes are in general. For integers $t > 1$, $j = 1, 2, \ldots, t$, and $i = 1, 2, \ldots, q^{t-1}$, we define a function called gap as follows:

$$\text{gap}(i, t, q) = \sum_{j=1}^{t} \left\lceil \frac{i}{q^{j-1}} \right\rceil - t. \quad (7)$$

**Theorem 5**: For prime power q, integer $t > 1$, let i, r, and p be integers such that $i = 1, 2, \ldots, q^{t-1}$, $r = 1, 2, \ldots, q$, and $p = q^t - iq + r + 1$. The 2-generator QT $[p(q^t-1)/(q-1), 2t, (p-1)q^{t-1}]_q$ code generated by the generator matrix given in (4) and (6) meets the Griesmer bound with a gap given by gap(i, t, q).

Proof: By the Griesmer bound, the length n of a code of dimension $k = 2t$, and distance $d = (p-1)q^{t-1}$ satisfies

$$n \geq \sum_{j=0}^{k-1} \left\lceil \frac{d}{q^j} \right\rceil = \sum_{j=0}^{2t-1} \left\lceil \frac{(p-1)q^{t-1}}{q^j} \right\rceil,$$

$$n \geq \sum_{j=0}^{t-1} \left\lceil \frac{(p-1)q^{t-1}}{q^j} \right\rceil + \sum_{j=t}^{2t-1} \left\lceil \frac{(p-1)q^{t-1}}{q^j} \right\rceil,$$

$$n \geq (p-1)(q^{t-1} + q^{t-2} + \ldots + 1) + \sum_{j=1}^{t} \left\lceil \frac{(p-1)}{q^j} \right\rceil,$$

$$n \geq (p-1)\left(\frac{q^t-1}{q-1}\right) + \sum_{j=1}^{t} \left\lceil \frac{(q^t - iq + r)}{q^j} \right\rceil,$$

Since

$$\sum_{j=1}^{t} \left\lceil \frac{(q^t - iq + r)}{q^j} \right\rceil = \sum_{j=1}^{t} q^{t-j} - \sum_{j=1}^{t} \left(\left\lceil \frac{i}{q^{j-1}} \right\rceil - 1\right),$$

so we have

$$n \geq (p-1)(q^t-1)/(q-1) + (q^t-1)/(q-1) - gap(i,t,q),$$

or,

$$n \geq p(q^t-1)/(q-1) - gap(i,t,q).$$

So the code meets the Griesmer bound with a gap defined by (7). Q.E.D.

If $i = 1$, $gap(1, t, q) = 0$. So we have the following result:

**Corollary 6**: For prime power q and integer $t > 1$, let r, and p be integers such that $r = 1, 2, \ldots, q$, and $p = q^t - q + r + 1$. The $[p(q^t-1)/(q-1), 2t; (p-1)q^{t-1}]_q$ code constructed meets the Griesmer bound with equality, and thus is length-optimal.

As an example, let us consider $t = q = 3$. Table I lists the calculated results on the codes obtained for $p = 17, 18, \ldots, 28$. It shows that the code becomes better when p increases. In the table, gb denotes the Griesmer bound on the length. But maximum p for a 2-generator QT two-weight code in the family is $q^t$, and when $p = q^t+1$, a 2-generator QT simplex code is obtained.

TABLE I
EXAMPLE OF CODES

| p | d | n | gb | gap | i | R | q |
|---|---|---|----|-----|---|---|---|
| 17 | 144 | 221 | 217 | 4 | 4 | 1 | 3 |
| 18 | 153 | 234 | 230 | 4 | 4 | 2 | 3 |
| 19 | 162 | 247 | 243 | 4 | 4 | 3 | 3 |
| 20 | 171 | 260 | 258 | 2 | 3 | 1 | 3 |
| 21 | 180 | 273 | 271 | 2 | 3 | 2 | 3 |
| 22 | 189 | 286 | 284 | 2 | 3 | 3 | 3 |
| 23 | 198 | 299 | 298 | 1 | 2 | 1 | 3 |
| 24 | 207 | 312 | 311 | 1 | 2 | 2 | 3 |
| 25 | 216 | 325 | 324 | 1 | 2 | 3 | 3 |
| 26 | 225 | 338 | 338 | 0 | 1 | 1 | 3 |
| 27 | 234 | 351 | 351 | 0 | 1 | 2 | 3 |
| 28 | 243 | 364 | 364 | 0 | 1 | 3 | 3 |

*ACKNOWLEDGMENT*

*The author would like to thank the anonymous referees and the Associate Editor for their helpful comments and suggestions that significantly improved the presentation of the results.*

*Eric Z. Chen* was born in Zhejiang, China, in 1963. He received the M. S. degree in Computer Science, and the Ph. D. degree in Electrical Engineering from Southwest Jiaotong University, Chengdu, China, in 1986 and 1991, respectively.

Since 1994, he has been with the Department of Computer Science, Kristianstad University College, Kristianstad, Sweden. His main research interests include information and coding theory, applied combinatorics, graph theory.